\documentclass[twocolumn]{aastex61}
\usepackage{mathrsfs, mathtools}
\newcommand{\as}{\arcsec{}}

\begin{document}

\title{Variable Outer Disk Shadowing Around the Dipper Star RX\,J1604.3-2130\footnote{Based on observations performed with SPHERE/VLT under program IDs 099.C-0341(A), 097.C-0902(A) and 095.C-0693(A).}}
\correspondingauthor{Paola~Pinilla, Hubble Fellow}
\affiliation{Department of Astronomy/Steward Observatory, The University of Arizona, 933 North Cherry Avenue, Tucson, AZ 85721, USA}
\email{pinilla@email.arizona.edu}

\author{P.~Pinilla}
\affiliation{Department of Astronomy/Steward Observatory, The University of Arizona, 933 North Cherry Avenue, Tucson, AZ 85721, USA}

\author{M.~Benisty}
\affiliation{Unidad Mixta Internacional Franco-Chilena de Astronom\'{i}a (CNRS, UMI 3386), Departamento de Astronom\'{i}a, Universidad de Chile, Camino El Observatorio 1515, Las Condes, Santiago, Chile}
\affiliation{Univ. Grenoble Alpes, CNRS, IPAG, F-38000 Grenoble, France}

\author{J.~de Boer}
\affiliation{Leiden Observatory, Leiden University, P.O. Box 9513, 2300 RA Leiden, The Netherlands}

\author{C.~F.~Manara}
\affiliation{European Southern Observatory, Karl-Schwarzschild-Str. 2, D-85748 Garching, Germany}

\author{J.~Bouvier}
\affiliation{Univ. Grenoble Alpes, CNRS, IPAG, 38000 Grenoble, France}

\author{C.~Dominik}
\affiliation{Anton Pannekoek Institute for Astronomy, University of Amsterdam, Science Park 904,1098XH Amsterdam, The Netherlands}

\author{C.~Ginski}
\affiliation{Leiden Observatory, Leiden University, P.O. Box 9513, 2300 RA Leiden, The Netherlands}
\affiliation{Anton Pannekoek Institute for Astronomy, University of Amsterdam, Science Park 904,1098XH Amsterdam, The Netherlands}

\author{R.~A.~Loomis}
\affiliation{Harvard-Smithsonian Center for Astrophysics, Cambridge, MA 02138, USA}

\author{A.~Sicilia-Aguilar}
\affiliation{SUPA, School of Science and Engineering, University of Dundee, Nethergate, Dundee DD1 4HN, UK}

\begin{abstract}
 Low brightness dips have recently been observed in images of protoplanetary disks, and they are believed to be shadows by the inner disk. We present VLT/SPHERE polarimetric differential imaging of the transition disk around the dipper star RX\,J1604.3-2130. We gathered 11 epochs that cover a large temporal baseline, to search for variability over timescales of years, months, weeks, and days. Our observations unambiguously reveal two dips along an almost face-on narrow ring (with a width of $\sim$20\,au), and the location of the peak of this ring is at $\sim$65\,au. The ring lies inside the ring-like structure observed with ALMA, which peaks at $\sim$83\,au. This segregation can result from particle trapping in pressure bumps, potentially due to planet(s). We find that the dips are variable, both in morphology and in position. The eastern dip, at a position angle (PA) of $\sim$83.7$\pm$13.7$^\circ$, has an amplitude that varies between 40\% to 90\%, and its angular width varies from 10$^{\circ}$ to 34$^{\circ}$. The western dip, at a PA of $\sim$265.90$\pm$13.0$^\circ$, is more variable, with amplitude and width variations of 31\% to 95\% and 12$^{\circ}$ to 53$^{\circ}$, respectively. The separation between the dips is 178.3$^{\circ}\pm$14.5$^{\circ}$, corresponding to a large misalignment between the inner and outer disks, supporting the classification of J1604 as an aperiodic dipper. The variability indicates that the innermost regions are highly dynamic, possibly due to a massive companion or to a complex magnetic field topology.   
\end{abstract}


\section{Introduction}     \label{introduction}
In the last decade,  extreme adaptive optics and coronagraphic observations at optical and near-infrared wavelengths in combination with high angular resolution and sensitivity observations at millimeter wavelengths have opened a new window for our understanding of planet formation. Powerful telescopes such as the 
Atacama Large (sub-)Millimeter Array (ALMA) and the Spectro-Polarimetric High-contrast Exoplanet REsearch (SPHERE) at the Very Large Telescope (VLT) have provided unprecedented angular resolution down to few astronomical units in protoplanetary disks. This has allowed us to resolve regions in disks where planets can leave an imprint on their nascent environment.  These recent observations of protoplanetary disks unveiled an incredible variety of structures, such as cavities, rings, gaps, spiral arms, arcs, and low brightness dips, which are likely related to the diversity of physical processes that rule planet formation and disk evolution. 

One of the most exciting sets of protoplanetary disks to study the processes of planet formation are those hosting dust depleted cavities, the so-called transition disks. Most of the disks revealing asymmetric features in scattered light, such as spiral arms \citep[e.g.,][]{muto2012, benisty2015, stolker2016} and shadows \citep[e.g.,][]{avenhaus2014, benisty2017, canovas2017} are in fact transition disks. At long wavelengths, these disks usually show, as a main feature, a large dust depleted cavity surrounded by a ring-like (symmetric or asymmetric) structure \citep[e.g.,][]{casassus2013, perez2014, zhang2014, pinilla2018, marel2018}.  This suggests that the physical processes dominating the evolution of these disks may create different structures in small (micron-sized) and large (millimeter/centimeter-sized) particles, producing a diversity of morphologies at multiwavelength observations.

From these sets of observations, low surface brightness dips in the outer parts of disks ($>$10\,au) observed in scattered light are of particular interest because they are believed to result from shadowing by a misaligned inner disk with respect to the outer disk, when the inner and outer disks have different inclinations and position angles (also called warped or broken disks). These observations allow us to directly connect the resolved outer regions with the unresolved inner disk. Scattered light observations of some transition disks suggest a small to intermediate misalignment between the inner and the outer disks \citep[HD\,135344B, LkCa\,15, TW\,Hya, DoAr\,44, HD\,143006;][]{debes2016, oh2016, stolker2016, benisty2018, casassus2018}, while observations of other transition disks suggest a large  misalignment \citep[$\sim$70$^{\circ}$, HD\,142527 and HD100453;][]{marino2015, benisty2017}. In addition, some of these shadows, as in the case of HD\,135344B, are variable within timescales shorter than a week, suggesting a very dynamic and asymmetric inner region. Interestingly, warps have also been observed at the very late stages of planet formation, such as in the debris disks AU\,Mic and $\beta$\,Pic \citep[e.g.,][]{golimowski2006, dawson2011, bocaletti2015}. Recently, \cite{casassus2018} identified shadows in another transition disk, DoAr\,44, that are not only detected in scattered light, but also in millimeter-observations, suggesting that the shadows are effective in cooling millimeter-dust grains. Molecular line ALMA observations, that show clear deviations from Keplerian motion, also support the idea that some transition disks are warped \citep[e.g.,][]{rosenfeld2012, casassus2015, loomis2017, walsh2017}.

Warps are mainly identified in two regimes depending on how they propagate \citep{papaloizou1983}. On one hand, warps can propagate following the diffusion equation when disks are thin, in particular when $h/r\lesssim\alpha$, where $\alpha$ is the dimensionless viscosity parameter \citep{shakura1973}, $h$ is the disk scale height, and $r$ is the distance from the central star. On the other hand, when disks are thick and $h/r\gtrsim\alpha$, warps are expected to propagate as bending waves  \citep{papaloizou1995}. In both cases, if the sound crossing time is longer than the induced precession time, the disk can warp significantly \citep[e.g.,][]{papaloizou_t1995, nixon2012, facchini2013, nealon2015}.  For the origin of these warps in protoplanetary disks, different possibilities have been proposed, such as a misaligned stellar magnetic field with respect to the rotation axis of the star or/and the disk \citep[e.g.,][]{bouvier2007, romanova2013}; a misaligned circumbinary disk with respect to the binary orbital plane \citep[e.g.,][]{foucart2013};  and the interaction with a massive planet or a binary companion on an inclined, and perhaps eccentric, orbit \citep[e.g.,][]{xiang2013, lubow2016, martin2016, owen2017, facchini2018}. The last scenario is likely the case for the transition disk around HD\,142527 where a $0.1\,M_\odot$ companion in an eccentric orbit with an inclination of $\sim$125$^\circ$ within a close to face-on disk \citep{lacour2016} may explain most of the observed properties, including a large cavity, shadows, spiral arms, deviation from Keplerian velocity in the CO lines, and the  horseshoe-like structure at dust millimeter continuum emission \citep{price2018}.

In this paper, we present new SPHERE polarimetric differential imaging of the transition disk around the star RX\,J1604.3-2130 \footnote{SIMBAD name} (2MASS J16042165-2130284, hereafter J1604) at different wavelengths.  This disk is a member of the Upper Scorpius association  and it is located at a distance of 150.2$\pm$1.4\,pc \footnote{J1604 is close by and the parallax has a small relative uncertainty, which justifies why the uncertainty is taken as symmetric.} \citep{gaia2018}. The disk hosts one of the largest dust cavities observed at millimeter wavelengths \citep{mathews2012, zhang2014, dong2017}. Observations from ALMA indicate that J1604 is as bright as the most luminous disks in younger regions (e.g., in Taurus and Lupus) in the millimeter, with its dust concentrated in a ring-like structure, and a dust mass of $\sim$40-50\,$M_\oplus$ \citep{barenfeld2016, pinilla2018}. J1604 was previously observed with scattered light with HiCIAO at 1.6\,$\mu$m \citep{mayama2012} and ZIMPOL/SPHERE at 0.626\,$\mu$m \citep{pinilla2015}. Both of these observations reveal a ring-like structure in scattered light located at $\sim$0.4\as from the star with a single dip along the ring, located in the east. Interestingly, in these two observations obtained  3 years apart, the dip was detected at different position angles (PA): $\sim$85$^\circ$ (HiCIAO) and $\sim$46$^\circ$ (ZIMPOL). If the dip originated from the same shadowing structure, its very fast rotation is inconsistent with the local Keplerian velocity at the ring position \citep{pinilla2015}. \cite{mayama2012} also reported a marginal detection of a second dip in the west at a PA of 255$^\circ$. 

J1604 was identified as a dipper \citep{ansdell2016}, which are young stellar objects (YSOs) that exhibit light curves punctuated by recurrent (periodic or aperiodic) dimming events on timescales of a few days, for example, from the  Convection, Rotation and Planetary Transits satellite (CoRoT), \emph{Spitzer}, and \emph{Kepler\,2}. Dippers are fairly common among YSOs \citep[$\sim30-40\%$,][]{alencar2010, cody2014, bodman2017} and are thought to be systems seen at high inclination, such that the dimming events are due to patches of dusty material that repeatedly occult the star as they cross our line-of-sight \citep[e.g.][]{scaringi2016, schneider2018}. The prototype dipper, AA\,Tau, exhibits periodic eclipses every 8.2\,days \citep{bouvier1999} and  a model where the inner edge of the accretion disk is warped by its interaction with the inclined stellar magnetosphere successfully explains the light curve and the spectral variability \citep{bouvier2007}. Aperiodic dippers have also been interpreted as resulting from clumps of dusty material passing our line of sight to the star, by related or possibly different mechanisms, such as vortices and forming planetesimals \citep{ansdell2016}. J1604 is very interesting in this context because it shows aperiodic dimming events and it is one of the deepest flux dips among the known K2 dippers \citep{ansdell2016}. However, its outer disk is close to face-on  \citep{mathews2012, zhang2014}, and hence its dipper nature (usually in highly inclined disks) suggests a possible strong misalignment between the inner and outer disks. In addition, it hosts a large ($\sim$83\,au) and highly depleted dust and gas cavity \citep{dong2017}.  In addition,  J1604 also evidences variable near- and mid-infrared excess \citep{dahm2009}. 

The paper is organized as follows. In Sect.~\ref{obs}, we describe the observations and the data reduction from our SPHERE observations at different epochs. In Sect.~\ref{analysis}, we present the data analysis, including the inspection of the radial profile for each observation, the characterization of the dips, and the comparison with recent ALMA observations.  In Sect.~\ref{discussion}, we discuss these results in the context of different origins for the shadows and the potential connection of their variability with the dipper nature of J1604. In addition, we also discuss in this section the potential origin of the observed cavity and the evidence of particle trapping from multiwavelength observations.  Finally, in Sect.~\ref{conclusion}, we provide the conclusions of this work.

\section{Observations}     \label{obs}

\begin{figure*}
 \centering
   	\includegraphics[width=17.5cm]{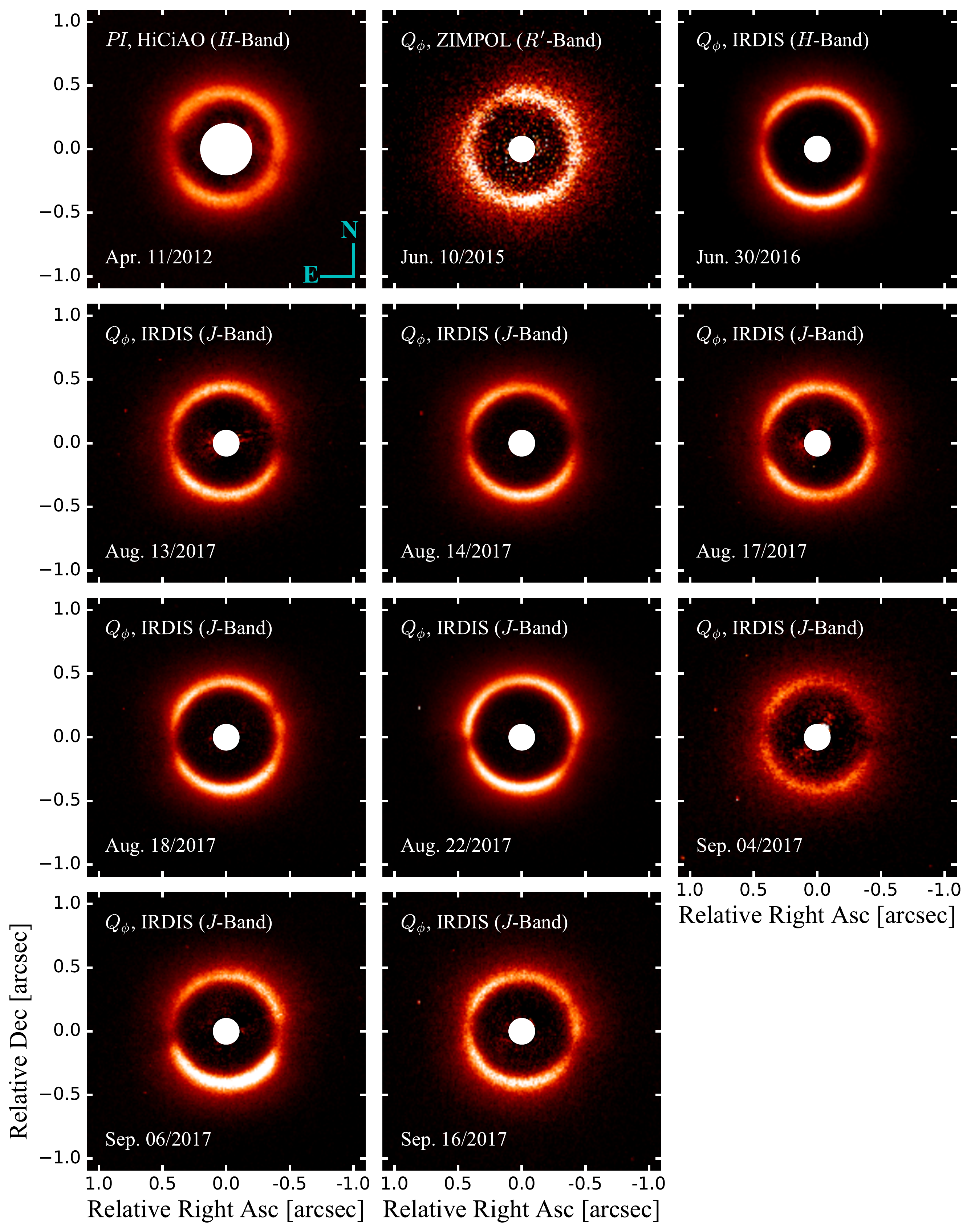}   
   \caption{Scattered light observations of the transition disk around J1604 (they are not scaled by $r^2$). The left upper panel corresponds to $H$-band  polarized intensity observations with HiCIAO reported by \cite{mayama2012}. The rest of the panels correspond to the Stokes parameter $Q_\phi$ obtained with VLT/SPHERE. The center upper panel corresponds to observations with ZIMPOL at $R'$-band. The right upper panel corresponds to observations with IRDIS at $H$-band, while the rest of the panels are IRDIS observations at $J$-band. In all the panels, the color scale is linear and in arbitrary units, and the dates are reported when the observations \emph{started}. This figure is available online as an animation.}
   \label{maps_J1604}
\end{figure*}

We obtained multiple epochs observations of J1604 at the Very Large Telescope located at Cerro Paranal, Chile, using the SPHERE instrument \citep{beuzit2008}, a high-contrast imager with an extreme adaptive optics system \citep{sauvage2014}. In this paper, we present new polarimetric observations covering nine epochs, obtained between 2016 June and 2017 September, in the near-infrared ($J-$ and $H-$bands) with the IRDIS instrument \citep{dohlen2008}. Our new data set is complemented by previously published visible ($R'$-band) polarimetric data obtained in 2015 June with ZIMPOL \citep[][]{pinilla2015} and near-infrared ($H$-band) polarimetric observations obtained with \emph{Subaru/}HiCIAO in 2012 April \citep{mayama2012}. This allows us to cover a large temporal baseline (2012-2017), and to search variability over timescales of years, months, and days.  In all our IRDIS observations, we used a 185\,mas diameter coronagraph (N\_ALC\_YJH\_S) to enhance the signal-to-noise ratio on the outer disk regions. The plate scale is 12.26\,mas per pixel.  We observed J1604 in excellent to good seeing conditions (between 0.5\arcsec{} and 1\arcsec{}).

The data reduction procedure is similar to the one reported in \citet{deboer2016} and is only very briefly described here. In polarimetric differential imaging, the stellar light is split into two orthogonal polarization states, and a half-wave plate (HWP) is set to four positions shifted by 22.5$^\circ$ to construct a set of linear Stokes images. The data is then reduced following the double difference method, from which one can derive the Stokes parameters $Q$ and $U$. If we assume single scattering events on the protoplanetary disk surface, the scattered light is linearly polarized in the azimuthal direction; therefore, for convenience, we describe the polarization vector field in polar coordinates with the $Q_\phi$, $U_\phi$ Stokes images \citep{schmid2006}. In this framework, the $Q_\phi$ image should contain all disk signals, while the $U_\phi$ image remains free of it. 

Fig.~\ref{maps_J1604} shows the Stokes parameter $Q_\phi$ of all of our SPHERE observations, including our previously published ZIMPOL image, and the $H$-band  polarized intensity observations with HiCIAO reported by \cite{mayama2012}. In all our new observations, we clearly detect two dark regions, hereafter referred to as dips or shadows, along the ring. These dips appear strongly variable with time, both in location and morphology.  

\section{Data Analysis}     \label{analysis}

\begin{figure*}
 \centering
   	\includegraphics[width=17.5cm]{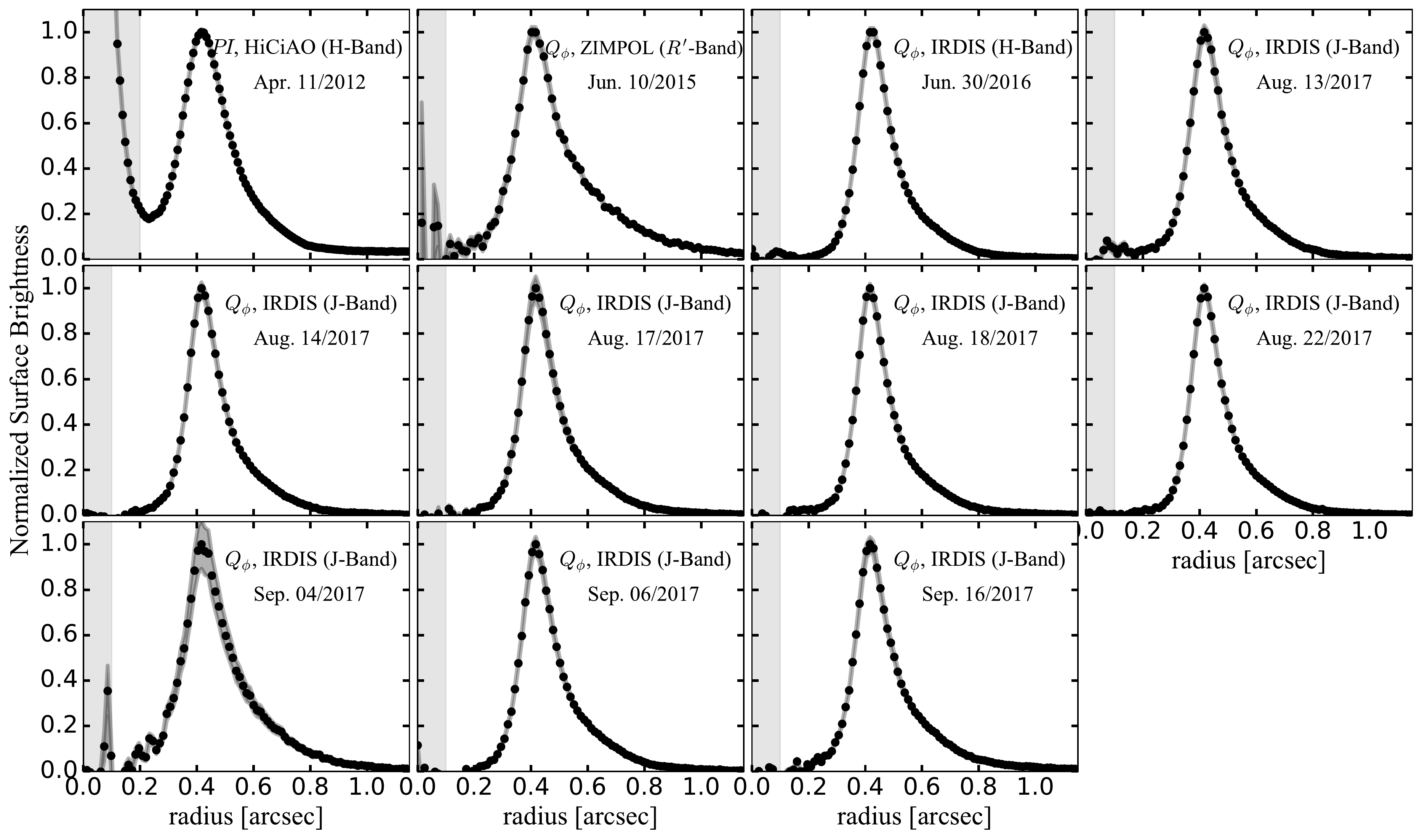}   
   \caption{Normalized radial profile of the surface brightness obtained after an azimuthal average. The uncertainty of the data correspond to the standard deviation in each radial bin, divided by the square root of the number of pixels in that bin. The shaded area corresponds to the coronagraph radius.}
   \label{radial_profiles_J1604}
\end{figure*}

\subsection{Radial profile} \label{sect:radial_profile}

\begin{table*}
\caption{Results from fitting the radial profile}
\label{table:all_rprofiles}
\centering   
\begin{tabular}{|c|ccc|c|}
\hline
\hline
&&\textbf{Lorentzian Fit}&&\textbf{Power-law Fit}\\
\textbf{Epoch}&$r_0$&$\gamma$&FWHM&$\xi$\\
&[$\as$]&[$\as$]&[$\as$]&\\
\hline
Apr.\,11/2012&0.43&0.10&0.19&-3.92$\pm$0.09\\
Jun.\,10/2015&0.43&0.10&0.20&-3.52$\pm$0.03\\
Jun.\,30/2016&0.43&0.06&0.13&-5.61$\pm$0.04\\
Aug.\,13/2017&0.43&0.07&0.14&-5.14$\pm$0.05\\
Aug.\,14/2017&0.43&0.07&0.13&-5.25$\pm$0.07\\
Aug.\,17/2017&0.43&0.06&0.12&-5.36$\pm$0.06\\
Aug.\,18/2017&0.43&0.07&0.14&-4.98$\pm$0.05\\
Aug.\,22/2017&0.43&0.06&0.12&-5.47$\pm$0.06\\
Sep.\,04/2017&0.43&0.09&0.18&-4.21$\pm$0.06\\
Sep.\,06/2017&0.43&0.07&0.13&-5.11$\pm$0.07\\
Sep.\,16/2017&0.43&0.07&0.14&-4.63$\pm$0.03\\
\hline
\hline
\end{tabular} 
\tablecomments{The uncertainties of the Lorentzian fit are omitted since they are negligible compared to the mean value.} 
\end{table*}

Figure~\ref{radial_profiles_J1604} shows the normalized radial profile of the surface brightness obtained after an azimuthal average. The uncertainty of the data corresponds to the standard deviation in each radial bin of the $Q_\phi$ images, divided by the square root of the number of pixels in the bin. We find a narrow bell-shaped curve, with a tail extending at larger radii. We note that this extended tail is not an effect of the instrumental point spread function (PSF). As a test, we convolved synthetic ring models, of various widths, with the PSF of each epoch and verified that this procedure cannot reproduce an extended tail beyond the peak as observed. Similarly, we also find that the PSF does not affect the dips' morphology, which is discussed in Sect.~\ref{sect:characterization_dips}. The details of the PSF shape for each epoch are summarized in Table~\ref{table:observing_log} in the Appendix. 

To quantify the location of the peak and the width of the ring, we fitted the radial profile of the normalized surface brightness of each epoch using a Markov chain Monte Carlo (MCMC) method. Considering the typical shape of the radial profile, we used a Lorentzian prescription, following:

\begin{equation}
A_L\times\left(\frac{\gamma}{(r-r_0)^{2}+\gamma^2}\right).
\label{eq:radial_Lorentzian}
\end{equation}

The model has three free parameters ([$A_L, r_0, \gamma$]), the amplitude, the location of the peak, and the half width at half maximum (HWHM), respectively. 
To perform the fit, we use {\it emcee} \citep{foreman2013}, which allows us to efficiently sample the parameter space to derive the maximum likelihood result for each model. The parameter space explored by the Markov chain for the location of the peak and the width are: $r_0\in[0.2\as, 0.6\as]$ and $\gamma\in[0.0\as, 0.2\as]$, with uniform prior probability distributions. The Markov chain sampled the parameter space for 1000 steps, with 100 walkers for each epoch. 
Table~\ref{table:all_rprofiles} summarizes the results from the fits for $r_0$, and $\gamma$, and includes the full width at half maximum (FWHM) too. The uncertainties from the MCMC fit are omitted since they are negligible compare to the mean value, with values of the order of $10^{-5}$ to $10^{-4}$ in all cases. 

\begin{figure*}
 \centering
   	\includegraphics[width=17.5cm]{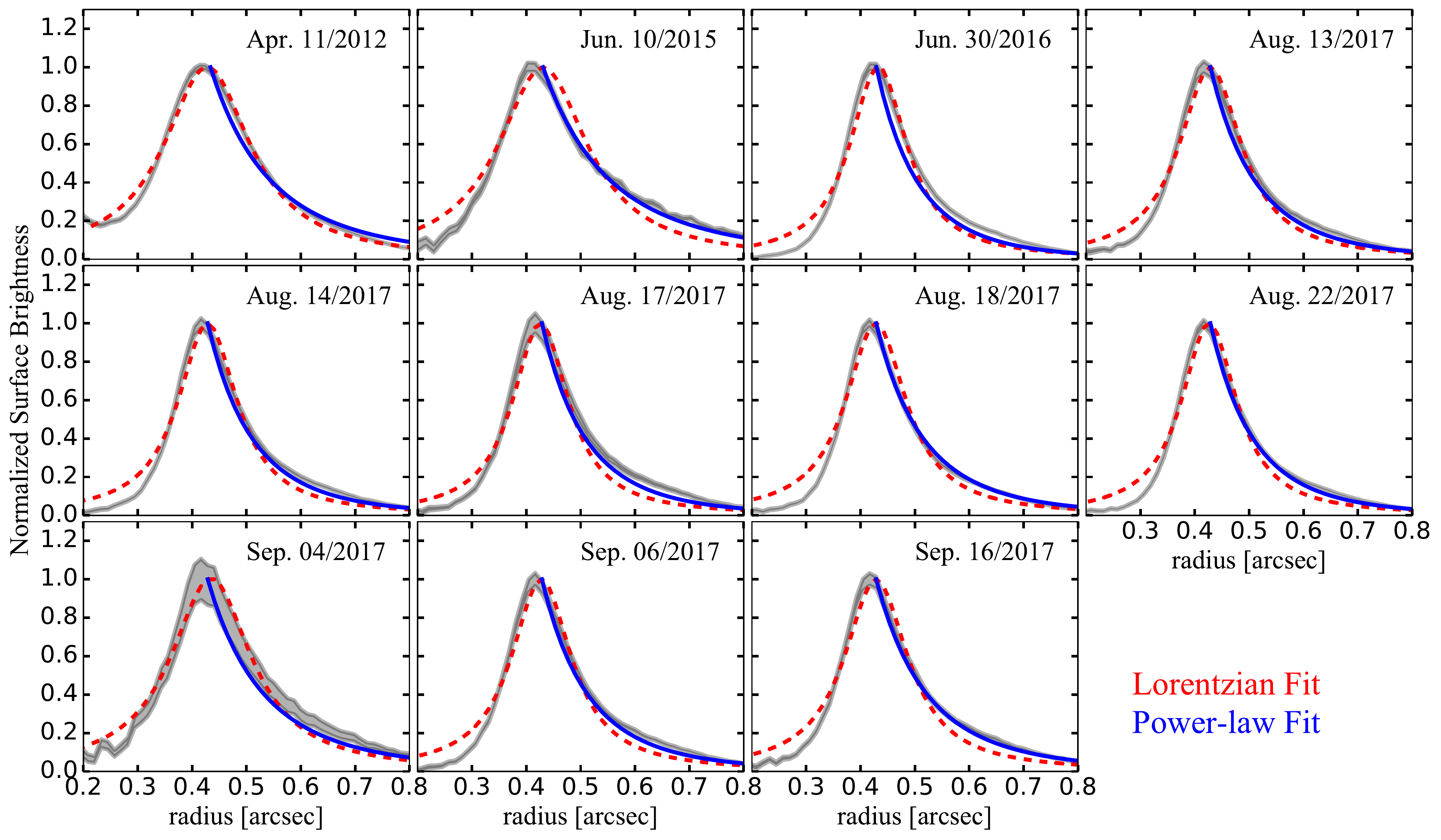}   
   \caption{Lorentzian MCMC best fit of the radial profile overlap with the uncertainty of the data profile for each epoch as shown in Fig.~\ref{radial_profiles_J1604}. Notice that the radial range changes from Fig.~\ref{radial_profiles_J1604}, to emphasize the ring shape and the fitting. In addition, we also overlay the power-law fit using non-linear least squares analysis and considering the data points beyond the peak.}
   \label{radial_fit}
\end{figure*}

\begin{table*}
\caption{Properties Of The Dips}
\label{table:all_dips}
\centering   
\begin{tabular}{|c|c|c|c|c|c|c|c|c|}
\hline
\hline       
\textbf{Epoch}&
min$_E$ [$^\circ$]&
min$_W$ [$^\circ$]&
min$_E$ [$^\circ$]&
min$_W$ [$^\circ$]&
A$_E$&
A$_W$&
$\sigma_E$ [$^\circ$]&
$\sigma_W$ [$^\circ$]\\
&
(image)&
(image)&
(fit)&
(fit)&
(fit)&
(fit)&
(fit)&
(fit)\\
\hline
Apr.\,11/2012&81.6&250.3&92.1$_{-1.8}^{+1.8}$&247.8$_{-7.9}^{+7.2}$&0.52$_{-0.06}^{+0.05}$&0.19$_{-0.06}^{+0.16}$&22.3$_{-2.7}^{+4.0}$&52.5$_{-2.7}^{+3.6}$\\
Jun.\,10/2015&42.2&---&44.9$_{-3.3}^{+4.7}$&---&0.25$_{-0.04}^{+0.04}$&---&28.2$_{-6.4}^{+8.8}$&---\\
Jun.\,30/2016&78.8&258.8&82.4$_{-2.0}^{+1.9}$&257.2$_{-1.3}^{+1.4}$&0.51$_{-0.06}^{+0.07}$&0.57$_{-0.05}^{+0.04}$&24.3$_{-3.9}^{+5.8}$&16.4$_{-1.3}^{+1.6}$\\
Aug.\,13/2017&92.8&267.2&83.1$_{-2.4}^{+2.4}$&274.4$_{-1.2}^{+1.3}$&0.50$_{-0.08}^{+0.14}$&0.67$_{-0.04}^{+0.04}$&31.4$_{-6.3}^{+10.3}$&21.9$_{-1.7}^{+2.1}$\\
Aug.\,14/2017&87.2&275.6&89.3$_{-1.6}^{+1.5}$&281.6$_{-1.3}^{+1.4}$&0.55$_{-0.05}^{+0.05}$&0.77$_{-0.04}^{+0.04}$&19.0$_{-2.2}^{+3.0}$&26.2$_{-2.4}^{+3.0}$\\
Aug.\,17/2017&90.0&270.0&90.7$_{-1.7}^{+1.6}$&280.5$_{-3.0}^{+2.8}$&0.52$_{-0.05}^{+0.05}$&0.31$_{-0.04}^{+0.04}$&17.3$_{-2.0}^{+2.7}$&20.7$_{-3.2}^{+4.1}$\\
Aug.\,18/2017&92.8&300.9&95.3$_{-1.4}^{+1.4}$&282.3$_{-2.4}^{+2.4}$&0.59$_{-0.05}^{+0.05}$&0.64$_{-0.15}^{+0.20}$&16.6$_{-1.6}^{+2.2}$&52.8$_{-10.6}^{+12.1}$\\
Aug.\,22/2017&98.4&247.5&98.5$_{-8.4}^{+1.5}$&249.7$_{-1.5}^{+1.6}$&0.44$_{-0.07}^{+0.10}$&0.47$_{-0.04}^{+0.04}$&9.7$_{-1.8}^{+3.5}$&15.5$_{-1.3}^{+1.5}$\\
Sep.\,04/2017&87.2&267.2&83.8$_{-1.4}^{+1.5}$&280.7$_{-2.6}^{+2.0}$&0.47$_{-0.06}^{+0.05}$&{\bf 0.78$_{-0.09}^{+0.10}$}&11.8$_{-1.5}^{+2.0}$&38.5$_{-6.2}^{+6.6}$\\
&&&&{\bf 268.0$_{-2.6}^{+2.6}$}&&0.58$_{-0.23}^{+0.15}$&&{\bf 16.3$_{-6.0}^{+8.3}$}\\
Sep.\,06/2017&92.8&264.4&77.2$_{-1.6}^{+1.4}$&288.7$_{-2.2}^{+2.1}$&0.90$_{-0.09}^{+0.06}$&{\bf 0.95$_{-0.08}^{+0.04}$}&33.8$_{-3.1}^{+3.2}$&54.9$_{-4.9}^{+3.8}$\\
&&&&{\bf 266.4$_{-2.1}^{+2.2}$}&&0.62$_{-0.23}^{+0.18}$&&{\bf 18.4$_{-5.0}^{+6.6}$}\\
Sep.\,16/2017&87.2&247.5&92.1$_{-2.1}^{+1.9}$&250.7$_{-2.2}^{+2.7}$&0.40$_{-0.05}^{+0.05}$&0.45$_{-0.07}^{+0.30}$&15.9$_{-2.1}^{+3.76}$&12.1$_{-2.2}^{+6.4}$\\
\hline
\hline
\end{tabular}   
\tablecomments{Best parameters from fitting a Lorentzian profile (Eq.~\ref{eq:radial_Lorentzian}), which are obtained for a range of PA of 96 to 350$^\circ$. For the epochs of Sep.\,04 and Sep.\,06, we report a second fit result considering a PA range from 236 to 300$^\circ$. For these two epochs, the values indicated in bold are the ones considered in the analysis. In addition, we report the angle at which the minimum is obtained from the image (i.e., dashed lines in Fig.~\ref{polar_plots})}
\end{table*}

According to our model fit, the peak of the ring is 0.43\as\,($\sim$65\,au) in all epochs. The FWHM varies from 0.12\as\,($\sim$18\,au; Aug.\,22/2017) to 0.18\as\,($\sim$27\,au; Sep.\,04/2017), in our $J$-band data. The epoch on Sep.\,04/2017 has the highest uncertainty from the IRDIS observations (Fig.~\ref{radial_fit}), and if we neglect this epoch, the width is almost constant and only varies from 0.12\as to 0.14\as. 
The FWHM of the ZIMPOL and HiCIAO data are 0.19\as\,and 0.20\as, respectively; but these observations have low signal-to-noise ratios compared to the IRDIS observations. 

To better quantify the extended tail of the ring, we performed a power-law fit to the surface brightness beyond the peak (i.e., taking the data from 0.43\as outward), such that it is proportional to $r^{\xi}$.  For this fit, we simply perform a nonlinear least squares analysis. The results for $\xi$ are also summarized in Table~\ref{table:all_rprofiles}, and the best models are overplotted in Fig.~\ref{radial_fit}.  The values of $\xi$ vary from -3.52 (ZIMPOL epoch on Jun.\,10/2015) to -5.61 (IRDIS epoch at H-band on Jun.\,30/2016). This power-law index can provide information about the vertical scale height of the disk, and its flaring index. 

Assuming a single scattering approximation, that is, the scattering of the starlight happens where the optical depth is unity,  the surface brightness of the disk is determined by the disk scale height (especially the scale height index $\beta$, such that $h(r)\propto r^\beta$, assuming hydrostatic equilibrium and vertically isothermal disks), and the radial profile of the surface density of the dust grains. For a disk with $h/r$ constant (i.e., a wedge-shaped disk), the surface brightness of the scattered light is expected to scale with radius as $\propto r^{-3}$, while for flared disks it is expected to scale as $\propto r^{\beta-3}$ \citep[e.g.,][]{whitney1992, dullemond2001, dong2012}.  The range of the slope that we found for J1604 is within the values obtained for several other HAeBe disks \citep{fukagawa2010}, which contradicts the flaring nature of several of these  disks. However, as suggested by \cite{fujiwara2006} and \cite{grady2007}, self-shadowing from, for example, the ring itself or a vertically thick inner wall, can create a steeper decrease of the surface brightness. \cite{dong2015} showed that indeed, the radial dependence of the surface brightness changes from a shallower $r^{-1.8}$ to a steeper $r^{-2.5}$ due to self-shadowing. 

Another way to explain steep power-law values as the ones we find is to consider a depletion of small grains. \cite{pohl2017} showed that the radial dependence of the surface brightness due to a large reduction in the distribution of small grains can change from $\sim r^{-2}$ to $\sim r^{-5}$. These models considered that grains are trapped in pressure maxima due to a planet carving a gap in a disk. Inside such a pressure bump, large grains accumulate, and  small grains are simultaneously generated due to fragmentation. At long times of evolution ($\sim$1-5\,Myr), most of the grains are located inside these pressure maxima \citep{pinilla2012}. In the case of J1604, both self-shadowing and a steep depletion of small grains beyond the ring can contribute to the values that we find for $\xi$. However, these values appear to change very quickly, suggesting rapid variations of the flaring, of the self-shadowing, or/and of the distribution of small grains at large radii.

\begin{figure*}
 \centering
  \begin{tabular}{cc}   
   	\includegraphics[width=\columnwidth]{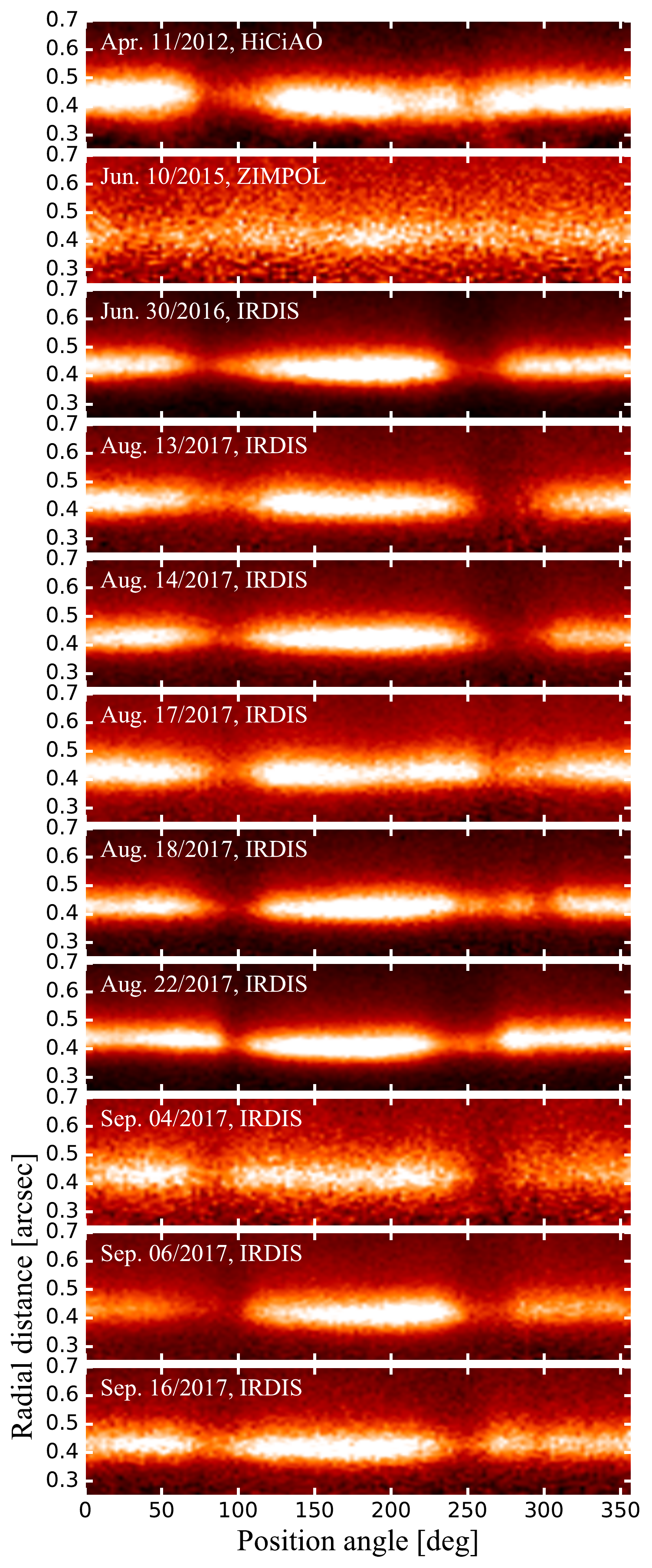}&
	\includegraphics[width=\columnwidth]{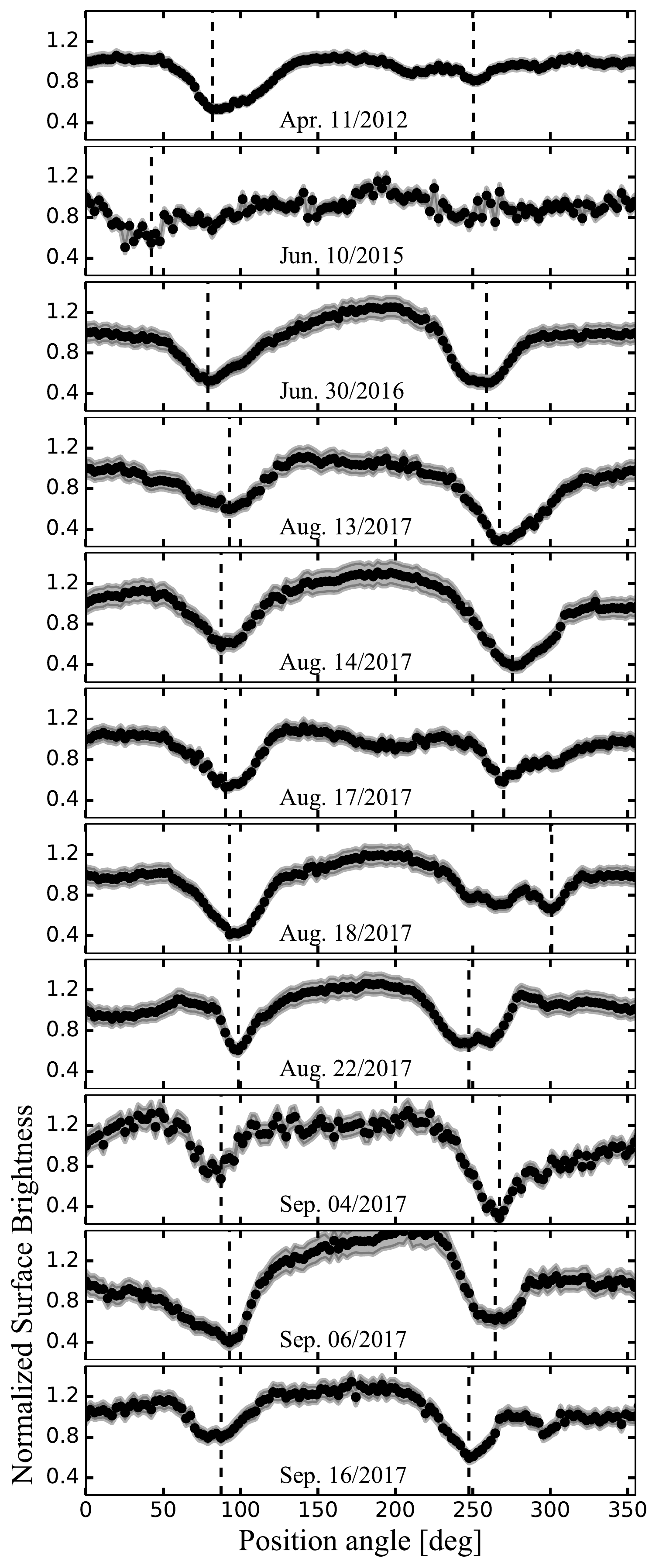}
   \end{tabular}
   \caption{Left panels: Radial mapping from 0.25-0.7\as of all the panels shown in Fig.~\ref{maps_J1604}. The color scale is linear and the maximum value taken in each case is 80\% of the maximum.  Right panel: azimuthal profile calculated from the mean values obtained between $[0.35-0.50]$\as. The shaded areas correspond to the uncertainty of the data and come from the standard deviation in the radial and azimuth divided by the square root of the number of pixels. The data is normalized to the value at zero degrees in each case. The dashed lines correspond to the minimum value obtained from the image between 0 and 150 degrees, and  200 to 350 degrees. For the ZIMPOL data only one minimum is shown.}
   \label{polar_plots}
\end{figure*}

\subsection{Characterization of the dips} \label{sect:characterization_dips}

Figure~\ref{polar_plots} shows the radial mapping from 0.25\as-0.7\as of all the images shown in Fig.~\ref{maps_J1604} (note that the images are not deprojected). The color scale is linear and the maximum value taken in each case is 80\% of the peak. In addition, the azimuthal profile is calculated for each epoch after radially averaging between $[0.35\as-0.50\as]$. The uncertainty (shaded areas) are the standard deviation of the data in the radial and azimuthal bins (i.e., $\sqrt{\sigma_{\rm{radial}}^2+\sigma_{\rm{azimuthal}}^2}$, where $\sigma_{\rm{radial}}$ and $\sigma_{\rm{azimuthal}}$ are the standard deviation of the data in the radial and azimuthal bins in the $Q_\phi$ images, respectively) divided by the square root of the total number of pixels within the ring. The data is normalized to the value at zero degrees in each case, which is our reference value to quantify the amplitude of these dips in the analysis that follows. We also test the following analysis normalizing to the peak of the data, which does not change the results. 

\subsubsection{Gaussian model}
To quantify the morphology of the dips and their variability, we performed a Gaussian fitting to each dip, similarly to the procedure followed with the radial profiles (Sect.~\ref{sect:radial_profile}). This means that we use {\it emcee}, and fit each dip independently with:

\begin{equation}
-A_{E, W}\times\left(\exp(-(\rm{PA}-\rm{min}_{E,W})^2/2\sigma_{E,W}^{2})+c_{E,W}\right)
\label{eq:radial_Gaussian2}
\end{equation}

\noindent The free parameters, $A_{E, W}$, $\rm{min}$$_{E,W}$, $\sigma_{E,W}$, and $c_{E,W}$, correspond to the amplitude, location of the minimum, Gaussian standard deviation, and shift of the Gaussian peak, respectively,  where $E, W$ indicates east or west for each dip. To fit each dip, we consider the data from $28^{\circ}$ to $150^{\circ}$ and from $196^{\circ}$ to $350^{\circ}$ for the eastern and western dip, respectively. For the ZIMPOL epoch, we only perform a fit for the eastern dip, since the western dip is not clearly detected.

\begin{figure}
 \centering   
   	\includegraphics[width=\columnwidth]{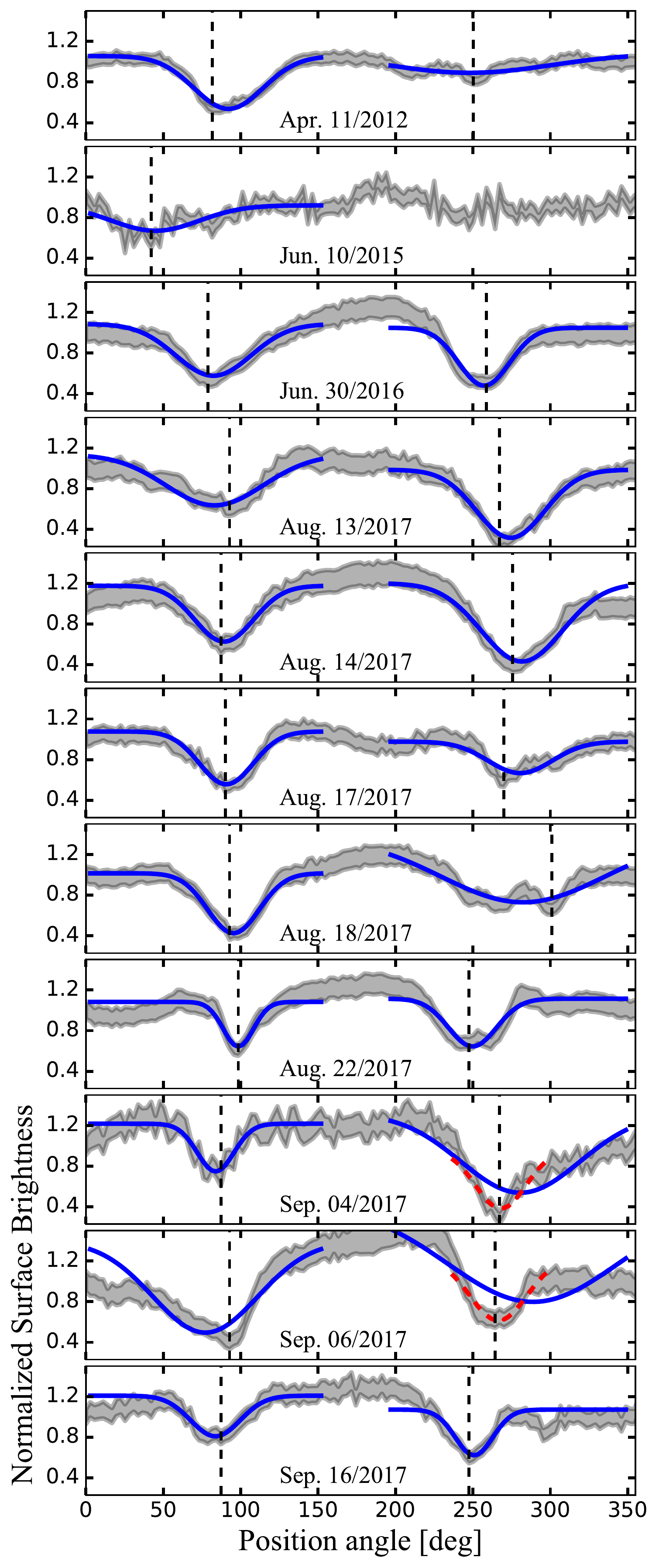} 
   \caption{Left panels: Gaussian MCMC best fit of each dip overlap with the uncertainty of the azimuthal profile for each epoch as shown in Fig.~\ref{polar_plots}. The fit shown in dashed red lines of the epochs Sep. 04 and Sep. 06/2017 corresponds to a Gaussian fit, but the PA range is taken from 236 to 300$^\circ$ for fitting the western dip of the epochs, instead of 196-350$^\circ$ as for the rest of the fits.}
   \label{dips_fits} 
\end{figure}

The parameter space explored by the Markov chain are: $A_{E, W}\in[0.1, 1.0]$,  $\rm{min}$$_{E}\in[28^{\circ}, 150^{\circ}]$, $\rm{min}$$_{W}\in[196^{\circ}, 350^{\circ}]$, $\sigma_{E, W}\in[5^{\circ}, 80^{\circ}]$, and  $c_{E,W}\in[0^{\circ}, 100^{\circ}]$, with uniform prior probability distributions.  The Markov chain sample the parameter space for 2000 steps, with 100 walkers for each epoch.  The results of the MCMC fit  for the parameters that quantify the dip morphology (i.e., $A_{E, W}$, $\rm{min}$$_{E,W}$, $\sigma_{E,W}$) are summarized in Table~\ref{table:all_dips}, and shown in Fig.~\ref{dips_fits}. 
We note that in most cases, our fitted $\rm{min}$$_{E,W}$ values are close to the minimum value of the radial profile obtained from the images, except in cases in which the dip shape is complex, either with multiple minima (e.g. western dip in epoch Aug.\,18/2017), or if there is a large difference in the levels of the radial profile on each side of the dips (e.g. eastern dip in epoch Sep.\,06/2017).
Indeed, for that reason, the fit is very poor for the western dip from the epochs of Sep.\,04 and Sep.\,06/2017. For these two epochs, we performed another fit by taking a narrower range for the PA from 236 to 300$^\circ$ (instead of 196-350$^\circ$), which provides a good match to the dip shape (see red dashed lines in Fig.~\ref{dips_fits}).

\begin{figure*}
 \centering
 \setlength{\tabcolsep}{0.1pt}
  \begin{tabular}{ccc}   
   	\includegraphics[width=6.1cm]{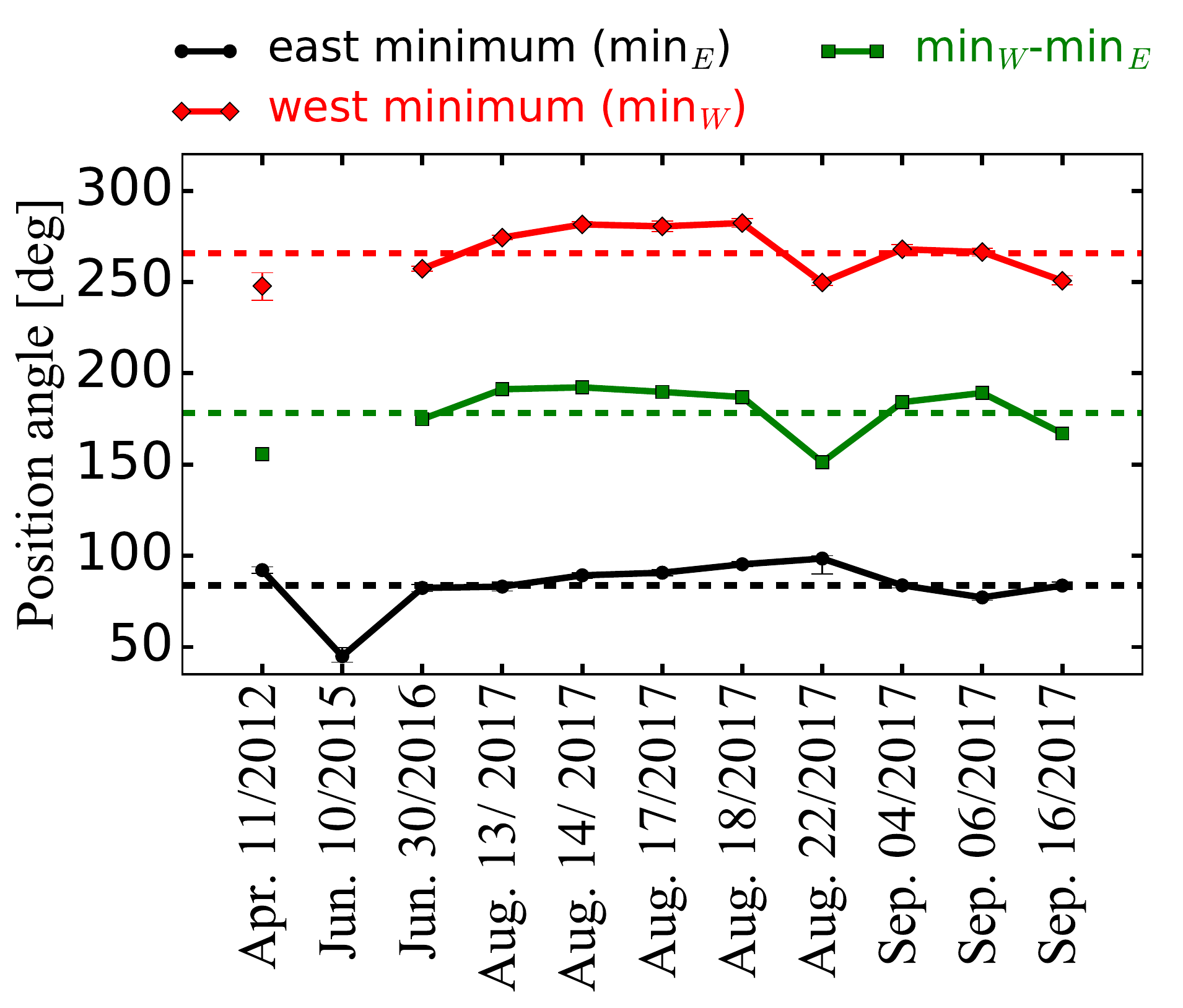}&
	\includegraphics[width=6.0cm]{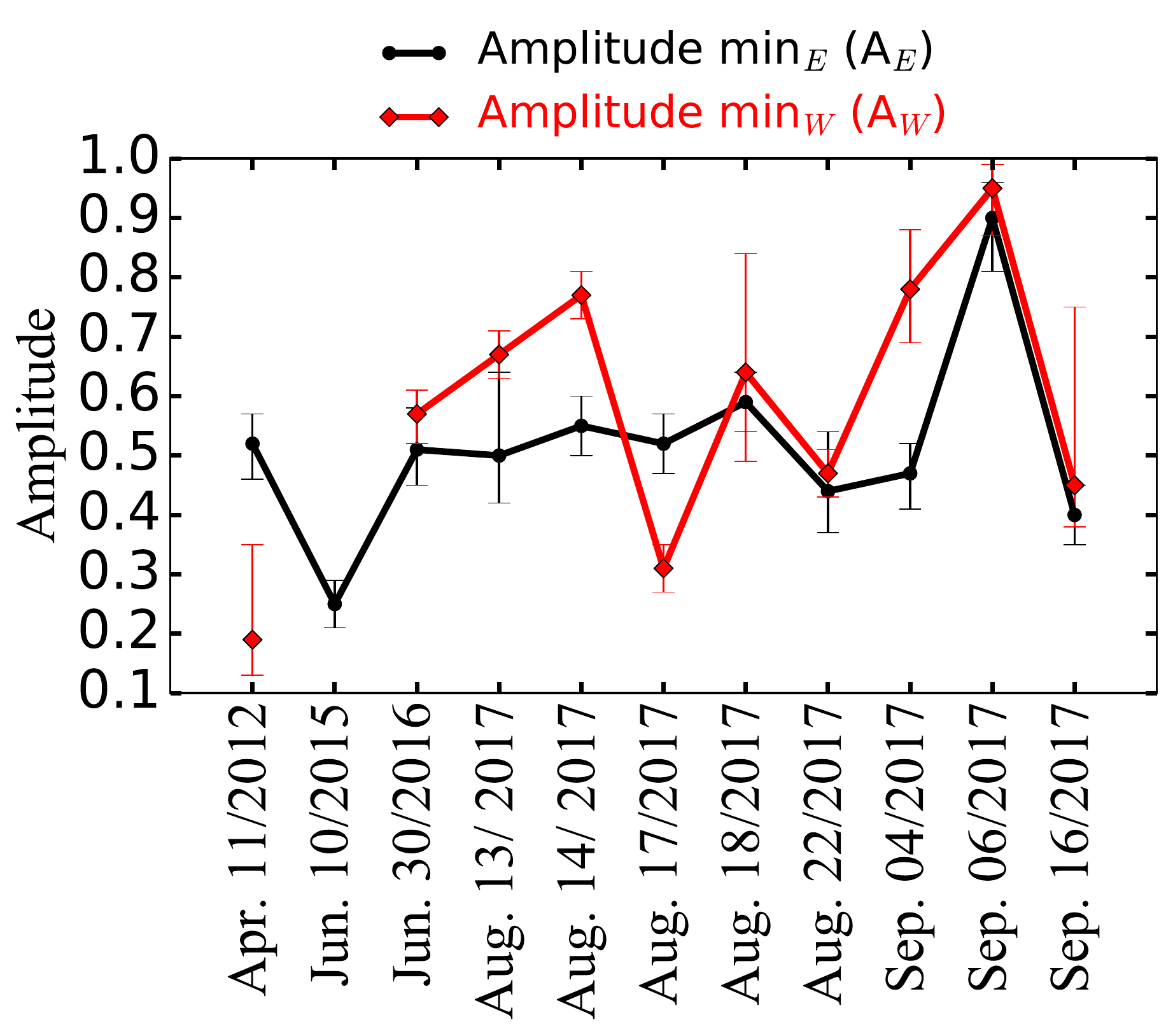}&
	\includegraphics[width=6.0cm]{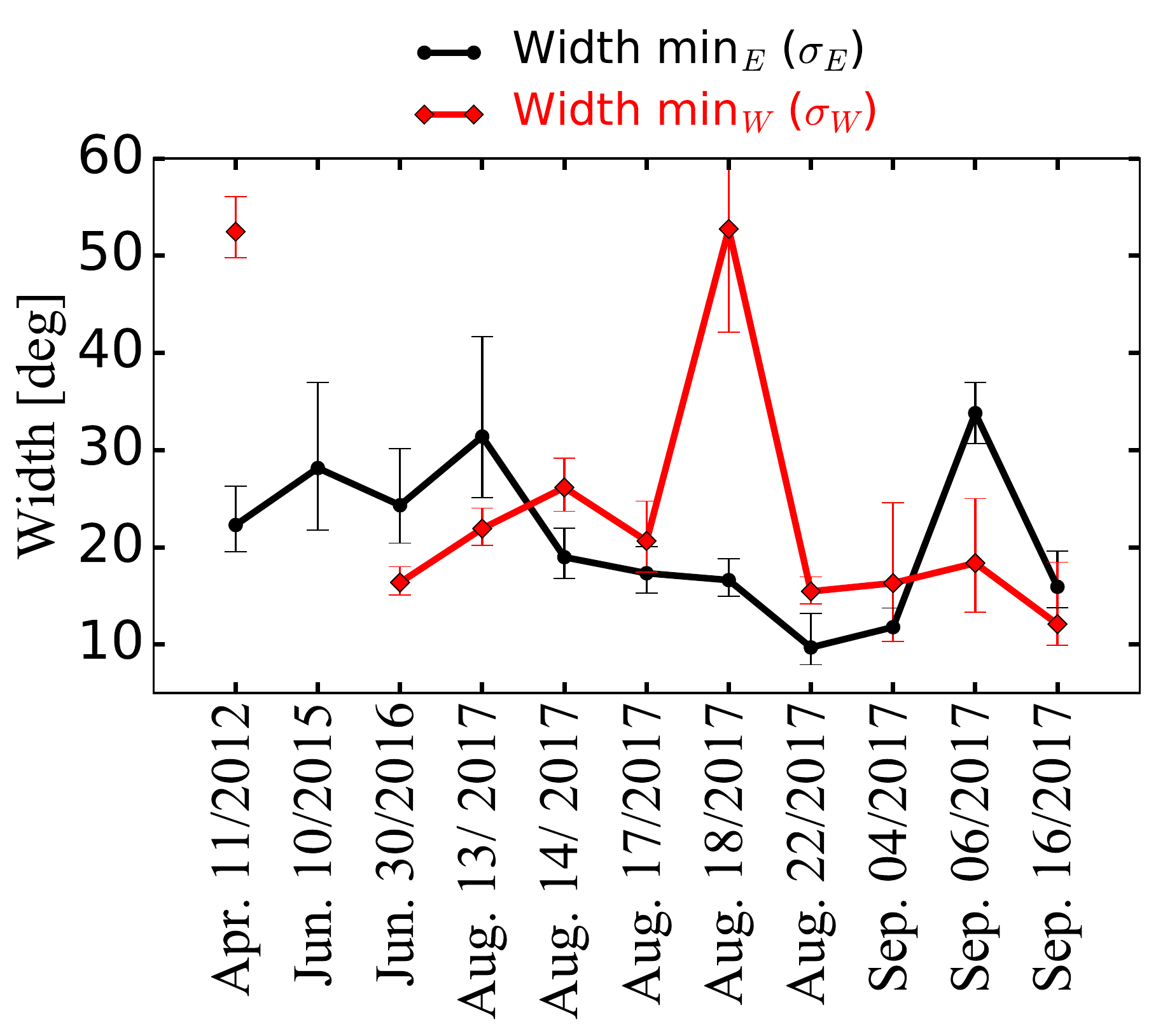}
   \end{tabular}
   \caption{Variations of the east and west minimum properties (i.e., results reported in Table~\ref{table:all_dips}). From left to right: location, amplitude, and width of each dip. In the left panel, the dashed lines correspond to the mean values including all epochs (i.e., $\overline{\rm{min}_E}=83.7^{\circ}\pm13.7^{\circ}$, $\overline{\rm{min}_W}=265.9^{\circ}\pm13.0^{\circ}$, and $\overline{\rm{min}_W-\rm{min}_E}=178.3^{\circ}\pm14.5^{\circ}$). For the western dip and the epochs of Sep.\,04 and Sep.\,06/2017, we take the location of the minimum and the width from the fit that takes the PA  range from 236 to 300$^\circ$, while for the amplitude, we take a PA range of  196-350$^\circ$, as for the rest of the fits. }
   \label{variations_from_fit}
\end{figure*}

\subsubsection{Variable morphology}
Fig.~\ref{variations_from_fit} shows the variations of the parameters (location, amplitude, and width of each dip). For epochs Sep.\,04 and Sep.\,06/2017, we use the minimum location and the width of the dip from the fit that assumes a narrow PA range, but the amplitude is taken from the fit with the large PA range, since it averages the amplitude before and after the dip and thus gives a more accurate value for this parameter.

\paragraph{Dip locations}
The location of the eastern dip varies from  $\sim$45$^\circ$ (ZIMPOL epoch) to $\sim$98$^\circ$ (epoch of Aug.\,22/2017), although the location of the minimum of the eastern dip from ZIMPOL epoch is an outlier in our sample (Fig~\ref{variations_from_fit}). The ZIMPOL data have low signal-to-noise ratios, but nonetheless we do not discard the existence of this dip at 45$^{\circ}$ in 2015. The mean value of the minimum location of the eastern dip is $\overline{\rm{min}_E}=83.7^{\circ}\pm13.7^{\circ}$. If we neglect the ZIMPOL epoch for this calculation, the mean value of the minimum location of the eastern dip is $\overline{\rm{min}_E}=87.6^{\circ}\pm6.3^{\circ}$, and the dip location varies within $\sim$16$^\circ$ (from 82 to 98$^\circ$). In the left panel of Fig.~\ref{variations_from_fit}, the dashed lines correspond to the mean values. The position of the western dip varies from $\sim$248$^\circ$ (HiCiAO epoch, Apr. 11/2012) to $\sim$282$^\circ$ (epoch of Aug.\,18/2017), although the location of the minimum of the western  dip from HiCiAO observations is not well constrained (see Table~\ref{table:all_dips} and Fig.~\ref{dips_fits}). If we neglect this value, the western dip varies within $\sim$32$^\circ$ (from 250 to 282$^\circ$). The mean value, including all the epochs in the sample is $\overline{\rm{min}_W}=265.9^{\circ}\pm13.0^{\circ}$, and the difference in location between the two dips is $\overline{\rm{min}_W-\rm{min}_E}=178.3^{\circ}\pm14.5^{\circ}$ (this last calculation neglects the ZIMPOL epoch). 

\paragraph{Dip amplitudes}
The amplitude of the eastern dip, derived from the near-infrared datasets, varies between 40\% (epoch of Sep.\,16/2017) to 90\% (epoch of Sep.\,06/2017). At the ZIMPOL epoch, the amplitude is 25\%, but this value might be affected by the low signal-to-noise ratio of the data. As for the location, the amplitude of the western dip appears to be more variable than the one of the eastern dip, with values ranging from from 31\% to 95\%.  In five epochs, the amplitude of the two dips appear to be similar. 

\paragraph{Dip widths}
The width of the eastern dip varies from $\sim$10$^{\circ}$ (Aug.\,22/2017)  to $\sim$34$^{\circ}$  (epoch of Sep.\,06/2017, also the one with the largest amplitude).  The width of the western dip varies from $\sim$12$^{\circ}$ (Sep.\,16/2017)  to $\sim$53$^{\circ}$ (Aug.\,18/2017).  The western dip of  epoch on Aug.\,18/2017 seems to be a composition of two different dips, and this may be the reason why the fit of this dip gives as a result of a very wide dip. There does not seem to be a relationship between the variation of the width of the two dips, and, on average, if we only consider the SPHERE data, the averaged width of the eastern and western dips are $\sim$20$^{\circ}$ and $\sim$22$^{\circ}$, respectively. %

In addition, we checked the radial profile outside of the dips, along the ring, from 140 to 200$^{\circ}$. In most of the cases, the surface brightness distribution is flat varying within 20\% of the reference value (at 0$^{\circ}$). In the epoch on Aug.\,17/2017, the surface brightness distribution decreases with PA, from values of 1.1 to 0.88. For the epoch on Sep.\,06/2017, the surface brightness distribution monotonically increases with PA, from values of 1.22 to 1.45. We note that this is the epoch with the largest amplitude for the two dips. 

\subsection{Comparison with ALMA observations} \label{sect:comparison_ALMA}
In \cite{pinilla2018}, we performed an analysis of the dust morphology of several transition disks, including J1604, that were observed with ALMA in the (sub-)\,millimeter regime. This analysis was done in the visibility plane to characterize the total flux, cavity size, and shape of the ring-like structure. Motivated by models of dust trapping in pressure bumps, we fitted a radially asymmetric Gaussian ring for the millimeter intensity, that is, a Gaussian ring whose inner and outer widths differ. For J1604, based on observations obtained with $\sim$0.26\as$\times$0.22\as~resolution, the inner and outer widths of the Gaussian from the best fit are $\sim$0.08\as~ and $\sim$0.14\as, respectively, while the Gaussian peaks at $\sim$0.55\as$\pm{0.01}$\as. The cavity size is well resolved while the width of the ring (0.22\as) remains unresolved.

\begin{figure}
 \centering
   	\includegraphics[width=\columnwidth]{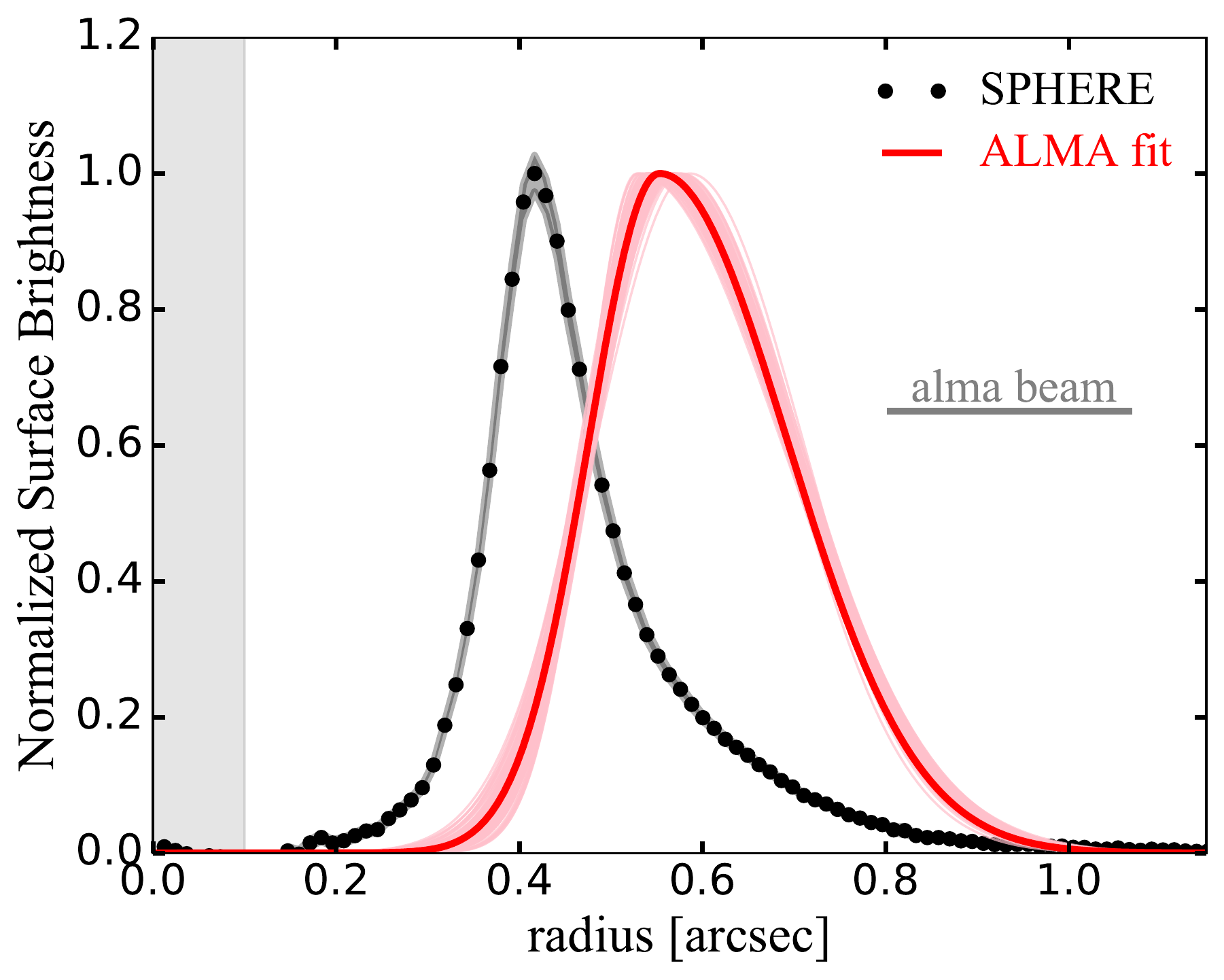}   
   \caption{Radial profile of the surface brightness, normalized to the peak of emission for SPHERE (black; Aug.\,14/2017) vs. ALMA \citep[red,][]{pinilla2018}. The horizontal gray line corresponds to the beam major axis from ALMA observations.}
   \label{SPHERE_ALMA}
\end{figure}

Figure~\ref{SPHERE_ALMA} shows the radial profile of the surface brightness, which is normalized to the peak of emission for SPHERE vs. ALMA. We randomly chose the epoch on Aug.\,14/2017 as a reference of our IRDIS observations. The ring observed in scattered light resides inside the cavity-observed with ALMA. It is expected that micron-sized particles also exists inside the ring observed at millimeter emission \citep{ovelar2013, pinilla2016}. However, shadowing from the ring itself can cause the ring observed in scattered light to be detected fully inside the ALMA ring \citep[see e.g., Fig~3 in][]{dullemond2010}. This shadowing effect supports the steepness of the surface brightness beyond the peak, as explained in Sect.~\ref{sect:radial_profile}. There is a significant separation between the two peaks (0.43 vs. 0.55\as, SPHERE vs. ALMA).  To compare with models of particle trapping by embedded massive planets as discussed in Sect.~\ref{discussion}, we calculate the ``wall" of the ring observed in scattered light \citep[$w_{\rm{SL}}$, defined as the radial location where the flux has increased by half from the minimum in the cavity and the peak of the ring,][]{ovelar2013}, we obtain that the wall location for the Aug.\,14/2017 epoch is $\sim$0.36\as. This implies that the ratio of $w_{\rm{SL}}$ and the peak of the millimeter emission is $\sim$0.65.

With a $\sim$0.25\as\,resolution, \cite{dong2017} analyzed $^{12}$CO, $^{13}$CO and C$^{18}$O J=2-1 line emission from ALMA observations of J1604. They concluded that their gas observations are consistent with a gas cavity that is smaller than the millimeter-dust cavity (with an upper limit for the inner radius at 0.10\as). From their thermo-chemical models, they suggested that the gas surface density smoothly  increases from 0.10\as\,to the peak of the millimeter emission and they exclude a sharp transition or double-drop models (i.e., models that assume two localized reductions) for the gas surface density. According to their results, the gas is depleted inside the cavity  by 2 to 4 orders of magnitude. Therefore, the ring observed in our scattered light observations lies in between the minimum of the gas surface density (inside 0.10\as)  and the peak of the millimeter emission (0.55\as).

\section{Discussion}     \label{discussion}

In this section, we discuss  the potential origin of the observed cavity and the evidence for particle trapping from multiwavelength observations. In addition, we also discuss different origins for the shadows and their variability, which can be potentially connected to the dipper nature of J1604.

\subsection{Origin of the Cavity and Evidence of Dust Trapping} \label{cavity_origin}
As a member of the Upper Scorpius OB association (one of the oldest star forming regions, 5-11\,Myr, that host protoplanetary disks), J1604 is an excellent target to investigate a critical stage when disk dissipation should be almost over \citep[e.g.,][]{williams2011}. Its spectroscopic signatures indicate very low accretion rates \citep[e.g.,][]{dahm2012}, recently confirmed with X-Shooter spectra (Manara et al. 2018, in preparation), with  $\log_{10}\dot{M_\star} [\,M_\odot\,\rm{yr}^{-1}$]=-10.54. This low value of the accretion rate may indicate a strongly gas depleted inner region, which is supported by the analysis of molecular gas \citep{dong2017}. Nonetheless, J1604 evidences variable NIR and MIR excess  \citep{dahm2009}, which can originate from an optically thick ring located at small ($\sim$0.1\, au) orbital radii. The potential existence of a dusty inner disk (also suggested from the shadows and the dimming events on the optical light curve), in addition to the low-mass accretion rate, and the comparison between the distribution of the gas and small/large grains (Sect~\ref{sect:comparison_ALMA}) can give constraints on the origin of the cavity. 

The spatial segregation between gas and small/large grains suggest particle trapping in pressure maxima, which naturally explains why this disk can remain quite massive in dust despite its old age. If a single planet is the primary cause of the cavity, the ratio of  the wall of the ring observed in scattered light ($w_{\rm{SL}}$), and the peak of the millimeter emission can hint at the mass of the embedded planet. \cite{ovelar2013} combined dust evolution models with hydrodynamical simulations of planet disk interaction to obtain the density distribution of different grain sizes when a massive planet in a circular orbit is embedded in the disk and filter dust grains of different sizes. These results, combined with radiative transfer predictions, allow us to infer the mass of the planet from observations of the ring-like structure of transition disks at different wavelengths \citep[see Fig.~8 in][]{ovelar2013}. The obtained value of the ratio of  the wall of the ring observed in scattered light of 0.65 suggests a planet mass of at least 4\,$M_{\rm{Jup}}$. 

Observations of CO and its isotopologues suggest that the location of this hypothetical planet or companion should be within $\sim$15\,au \citep[assuming a distance of 150.2\,pc,][]{gaia2018}, which corresponds to the location of the minimum of the gas surface density (0.10\as). This location, however, remains unresolved from ALMA observations. The mass and location of such hypothetical planet are below the upper limits on close companions derived from \cite{kraus2008} ($\sim$70\,$M_{\rm{Jup}}$ between 10-20\,au), \cite{ireland2011} ($\sim$83\,$M_{\rm{Jup}}$ within 45\,au), and \cite{canovas2017} (2-3 $M_{\rm{Jup}}$ from 22 to 115\,au). Nevertheless, the mass of this  potential planet seems to be too high for the system to maintain an inner disk at the age of Upper Sco. Such a massive planet will block most of the dust (of all sizes) at the outer edge of the planetary gap such that after $\sim$5\, Myr of evolution no dusty material would remain in the inner disk \citep{pinilla2016}. 

Alternatively, the cavity can form due to multiple planets,  which leads to wider and shallower cavities \citep{duffell2015}, and allow a flow of dust from the outer  to the inner disk.  This dust is expected to drift to the very inner regions ($\sim$1\,au) and can pile up at these small orbital radii because of their low drift velocities near the snow line \citep{pinilla2016}. Assuming that the disk effective temperature results from stellar irradiation and accretion, the location of the snow line is expected to be at $\sim$1\,au, considering a stellar luminosity of $\sim$0.58\,$L_\odot$,  a stellar temperature of 4500\,K, and a stellar mass of $\sim$1\,$M_\odot$ \citep{dahm2009, mathews2012}. However, in this scenario, the gas in the cavity would not be as depleted as suggested by the ALMA observations and  the accretion rate would not be that low. 

The possibility of a close binary is excluded from high-resolution optical spectroscopy, which does not show signs of a double-line spectroscopic binary \citep{dahm2012}. However, a massive companion in a wide eccentric orbit (as in HD\,142527) still remains as a possibility. In this case, the inner disk may be filled by streamers bridging it to the outer disk. When a companion is massive enough ($>$5\,$M_{\rm{Jup}}$, depending on the disk viscosity), the disk becomes eccentric and streamers are more efficient at transporting material \citep{ataiee2013}. As a result, accretion of material onto the planet and flows of material from the outer disk to the inner disk can be enhanced \citep{kley2006,ragusa2017}. In this case, an inner disk can be maintained for longer times of evolution, and the accretion is expected to be variable with time for both the central star and the companion \citep{munoz2016}. Hence, an eccentric planet could be a viable solution to allow the inner disk to be replenished and at the same time be consistent with the lower limit for the planet mass inferred from the dust segregation, as discussed above.

In the inner disk, where the gas density is depleted, kilometer-sized planetesimals would be completely decoupled from the gas and they would not experience fast inward radial drift. These planetesimals can potentially collide, recreating a belt of micron-sized particles in the inner disk. These micron-sized particles are expected to grow efficiently, unless they reside (or grow and drift) inside the snow line where fragmentation of silicates is efficient, keeping a sufficient amount of small-grains close to the star. This scenario would imply that the inner disk extent is very small (within the snow line, i.e.,$\sim$0.1-1\,au).

Other  possibilities for the formation of the cavity and particle trapping include photoevaporation and non-ideal magnetohydrodynamical (MHD) effects. On one hand, photoevaporation is consistent with the low accretion rate of J1604, and the detection of  the [OI] line (Manara et al. 2018, in preparation). \cite{ercolano2018} also demonstrated that  X-ray photoevaporation in a disk with a moderate gas depletion of carbon and oxygen can create cavities as big as 100\,au with a large range of accretion rates ($\dot{M_\star}\sim10^{-11}-10^{-8}\,M_\odot\,\rm{yr}^{-1}$). However, as pointed out by \cite{dong2017}, models of photoevaporation predict a sharp cavity edge in gas \citep[e.g., see Fig.\,1 in][]{ercolano2018,alexander2006}, which contradicts the results from the analysis of the CO observations from ALMA. 

Nonideal MHD effects, such as dead zones can also create cavities observable at different wavelengths \citep[e.g.,][]{flock2015, ruge2016} because particle trapping can occur at the outer edge of the dead zone where there is a bump in the gas density profile due to the change of accretion from the dead to the active MRI regions. This scenario, however, predicts that cavities at short and long wavelength should be of similar size \citep{pinilla2016a}. A solution to create spatial segregation, as observed in J1604,  is the inclusion of an MHD wind (possibly traced by the [OI] emission line; Manara et al. 2018, in preparation) to the dead zone models, which can create a large difference in the distribution of gas and small/large particles. In this case, the gas surface density inside the cavity can be depleted by several orders of magnitude and increases smoothly with radius \citep[see Fig. 6 in][]{pinilla2016a}, as suggested for J1604 by ALMA observations. 

\vspace{1cm}
\subsection{Shadows and Their Variability}
Dips in scattered light images have been interpreted  as shadowing from a misaligned inner disk \citep{marino2015, benisty2017}. Assuming an outer disk inclination and position angle of $i$=6$^\circ$ and PA=80$^\circ$ \citep{pinilla2018}, respectively,  and a disk aspect ratio of 0.1 for the scattering surface at the ring radius \citep{dong2017}, we find that the inner disk should be close to edge-on, leading to a misalignment between inner and outer disks of $\sim$70-90$^\circ$ \citep[see equations in][]{min2017}. A strongly inclined inner disk is consistent with the dipper activity of this object, by which dimming events can be caused by patches of dusty circumstellar material that repeatedly occult the star as they cross the line of sight.

The intersection of the planes of the inner and outer disks defines the PA of the shadows (assuming that they are razor-thin). Due to the finite scale height of the inner disk, the shadows appear as broad dark regions, and their widths can in principle constrain the scale height of the inner disk. If the relative orientation of the disks is fixed with time, the PA of the shadows should not vary. However, we find that it varies within $\sim\pm$14$^\circ$, from the estimate of the (local) minimum value of the surface brightness. It is possible that the PA variations are related to the variations of the dips' shapes and widths, which can in turn modify the location of the minimum surface brightness that we estimate (or PA). Such variations in the widths and PA of the shadows imply that they are not caused by a symmetric inner disk with a constant misalignment with respect to the outer disk.  Instead, it is likely that the inner disk is highly structured and asymmetric, and that its scale height varies with time, in very short time scales (within a day). In addition, the fact that the shadow properties and variability are very different for the eastern and western dips, also supports an asymmetric morphology of the inner disk. 
  
Both the fast dynamics and asymmetric morphology of the shadows are likely connected with the aperiodicity of the stellar dimming events (assumed to be due to an inner disk warp), which can change as much as $\sim$60\% also in time scales of few days in dipper objects \citep{ansdell2016, cody2018}. These variations likely originate from variations of the inner disk scale height of the order of 10\% or more \citep{mcginnis2015}. 
  
It is therefore not surprising that the width of the shadows observed in J1604 significantly vary on day to week timescales, as they directly reflect the intrinsic variations of the inner disk scale height on these timescales. Furthermore, no clear correlation is expected between the widths of the east and west shadows as long as the inner disk scale height varies on a timescale shorter than a few days, which appears to be the case for most dippers \citep{mcginnis2015}. The maximum amplitude of the dips, however, depends on the optical thickness of the inner disk warp, i.e., the dust properties. The seemingly correlated amplitudes of variability of the east and west shadows might mean that the dust properties (in particular, opacity) change on longer timescales than the inner disk warp shape.

A misaligned stellar magnetic field with respect to the rotation axis of the star can create a magnetically warped inner disk edge as in the case of the disk around AA\,Tau \citep{bouvier1999}. Such a warp occults AA\,Tau periodically, and accounts well for the spectral variability of this system \citep[e.g.,][]{bouvier1999, bouvier2003, bouvier2007, menard2003}. This is because AA\,Tau is a fully convective star that hosts a very strong (2-3\,kG) \emph{dipolar} magnetic field, which maximizes the star-disk interaction, and produces a fairly stable, though dynamical, inner disk warp. However, J1604 is a more massive star with a well-developed inner radiative core. Partly radiative  pre-main-sequence stars tend to exhibit weaker fields that are mostly octupolar \citep[e.g., V2129 Oph,][]{donati2011, gregory2012}. In this case, more complex accretion flows are expected \citep{alencar2012}, possibly leading to more unstable and aperiodic star-disk interactions. If the octupole dominates at the disk level, it is possible to have an asymmetric disk warp with a complex perturbation of the inner disk scale height as a function of azimuth. The observed variations of the shadows PA could therefore result from the varying vertical shape of the inner disk warp as a function of azimuth.
Thus, a complex magnetic field geometry, coupled with a relatively weak accretion rate, can produce the strong and irregular variability seen in the shadows of J1604. Nonetheless, there are currently no observational estimates of the magnetic field strength and topology for J1604, and whether a complex magnetic field topology can be the origin of the observed variability and shadows still has to be investigated with a dedicated spectro-polarimetric campaign.

Apart from a misaligned magnetic field, the presence of a yet-undetected inclined massive companion in the cavity of J1604 could be responsible for a large misalignment between inner and outer disks. Indeed, in the case of HD\,142527, there is strong evidence that a $0.1\,M_\odot$ companion in an eccentric orbit with an inclination of $\sim$125$^\circ$ \citep{lacour2016} is the cause of the misalignment of the inner disk with respect to the outer disk, which can also explain most of the observed properties of this disk \citep{price2018}. While a single massive planet or companion in a coplanar and circular orbit might not explain the spatial segregation of the gas and small/large grains (see Sect~\ref{cavity_origin}), we cannot exclude the presence of such an inclined and eccentric companion in the cavity of J1604. 

The variation of the shadows' properties might be to a large number of dust clumps, orbiting with a large range of inclinations, maybe due to planetesimal collision in the depleted inner regions. The effect of light travel time can create shadows with a large range of morphologies, from arc-shaped to spiral arms, depending on the disk scale height (flat vs. flared disk) and the disk inclination \citep{kama2016}. In the case of J1604, due to the narrow extent of the ring in the outer disk, the expected shadows by clumps in the inner disk would not look like spirals but instead, as localized dips within the ring of emission. In any case, if this mechanism is responsible for the shadows, these clumps must be very dynamic (changing position and morphology in day timescales) to explain the observed variability of the dips, and lead to a sufficient amount of small dust that would lead to a large radial optical depth. Detailed modeling to assess if this is possible is required.

Another possibility to explain the presence of shadows and the observed variability is to consider the effects of magnetohydrodynamic instabilities. The Parker instability that occurs when amplified magnetic fields (by disk dynamo) can escape from the disk due to magnetic buoyancy \citep{takasao2018}. As a result of angular momentum exchange mediated by magnetic fields, the velocity and density above the disk increases and magnetic fields can escape due to magnetic buoyancy. As a consequence, due to the MRI-driven turbulence and  eruptions of the magnetic field, the density near the disk surface can significantly fluctuate spatially and temporally. In addition, due to MRI turbulence, the upper disk layers that are magnetically supported can carry dust grains at high altitude, and lead to shadows with typical timescales from half to a tenth of the Keplerian period at the inner disk \citep{turner2010}, potentially explaining aperiodic dimming and shadowing events.

In all scenarios, if the shadows are steady over orbital timescales at the ring radius, and the cooling timescale is  comparably fast, they could lead to a decrease of the dust and gas temperature in the ring, and also appear as dips in the millimeter images. There is marginal evidence for shadows at millimeter emission from ALMA observations \citep{dong2017}, similarly to the case of DoAr\,44 \citep{casassus2018}. This aspect will be further investigated in Loomis et al (2018, in preparation).

\vspace{0.5cm}
\section{Conclusions}     \label{conclusion}
We present new VLT/SPHERE polarimetric differential imaging of the transition disk around the dipper star J1604. We gathered a total of 11 epochs of scattered light imaging that span days, weeks, months, and years (Fig~\ref{maps_J1604}). Our findings are:

\begin{enumerate}
\item All the scattered light epochs show a ring-like emission that peaks at $\sim$0.43\as\, from the star (Fig.~\ref{radial_profiles_J1604}).  The morphology of the outer tail of this ring rapidly changes with time (Fig.~\ref{radial_fit} and Table~\ref{table:all_rprofiles}). The width of this ring remains unresolved in our observations (as in the case of millimeter observations with ALMA), with a value of $\lesssim$0.13\as. This ring lies inside the cavity resolved at millimeter emission, which also shows a ring-like structure but peaking at 0.55\as (Fig.~\ref{SPHERE_ALMA}). This spatial segregation can be a natural result of particle trapping in pressure bumps, created by, for example, embedded planet(s).

\item In the case of a single massive planet being the origin of cavity, at least a 4\,$M_{\rm{Jup}}$ mass planet is required inside the cavity to create the observed segregation of small and large grains. However, such a planet cannot explain the gas surface density structure inferred from ALMA observations and the existence of an inner dust optically thick belt, at the age of Upper Sco,  to explain the (variable) NIR excess. Potential alternatives include the possibility of a dead zone and a MHD wind acting \emph{together} in the evolution of J1604.

\item We detect two clear dips of emission along the ring observed in scattered light (Figs.~\ref{radial_profiles_J1604} and~\ref{polar_plots}). Both dips are highly variable in amplitude and width (Fig.~\ref{variations_from_fit}). The western dip is in general more variable than the eastern dip. For the eastern dip, the amplitude varies from 40\% to 90\% and its width varies from 10$^{\circ}$ to 34$^{\circ}$. For the western dip, the amplitude varies from 31\% to 95\% and its width varies from 12$^{\circ}$ to 53$^{\circ}$. 
From the 11 epochs, 
the mean position of the dips are $\sim$83.7$\pm$13.7$^\circ$ and $\sim$265.9$\pm$13.0$^\circ$ for the eastern and western dip, respectively. The averaged separation between the dips is 178.3$^{\circ}$$\pm$14.5$^{\circ}$. 

\item Assuming that these dips are shadowing from a misaligned inner disk, we find that the misalignment between the inner and the outer disk is very large ($\sim$polar), and similar to the values found for the other two transition disks: HD\,142527 and HD\,100453. Current available observations do not provide constraints on what can be the origin of the warp, and it remains an open question if it is due to a companion in an highly inclined orbit (as for the case of HD\,142527), a  misaligned stellar magnetic field with respect to the rotation axis of the star (as for the case of AA\,Tau), or other alternatives, such as dusty asymmetric clumps from forming planetesimals. The variability of the morphology of the shadows, along with the rapid variations of the morphology of the ring tail, suggest that the innermost regions are highly dynamic and complex.

\item The misalignment between the inner and the outer disk reconciles the dipper activity of J1604 since dimming events in light curves are mainly observed in highly inclined disks. Future demographic studies are needed to test if close-to-face-on dippers show shadows in scattered light and vice versa. 

\end{enumerate}

Future VLTI, spectro-polarimetric campaign, high-resolution ALMA observations and simultaneous optical/IR light curves can help us to better characterize  the inner disk causing the dipping events and shadows and the stellar magnetic field. In addition, ALMA line observations can also provide potential variations of the Keplerian motion in the inner regions of J1604 from different molecular lines, which will help us to characterize the warp and the possible presence of massive companion(s) in the disk responsible for the observed cavity.  

\paragraph{Acknowledgments}
  \acknowledgments{We acknowledge the referee for a constructive report that helped improve the paper. The authors are thankful to A.~Natta, S.~Facchini, A.~Pohl, and M.~Ansdell for interesting discussions on J1604. We thank S.~Mayama for sharing the HiCIAO image. P.P. acknowledges support by NASA through Hubble Fellowship grant HST-HF2-51380.001-A awarded by the Space Telescope Science Institute, which is operated by the Association of Universities for Research in Astronomy, Inc., for NASA, under contract NAS 5-26555. M.B. acknowledges funding from ANR of France under contract number ANR-16-CE31-0013 (Planet Forming disks). C.F.M. acknowledges support through the ESO Fellowship. J.B. acknowledges funding from the European Research Council (ERC) under the European Union's Horizon 2020 research and innovation programme (grant agreement No 742095; {\it SPIDI}: Star-Planets-Inner Disk-Interactions). C.D. acknowledges funding from the Netherlands Organisation for Scientific Research (NWO) TOP-1 grant, project number 614.001.552. R.A.L. gratefully acknowledged funding from NRAO Student Observing Support.} 

\software{Numpy \citep{walt2011}, SciPy \citep{jones2001}, Matplotlib \citep{hunter2007}, emcee \citep{foreman2013}.}

\appendix
\section{Observational Notes}
Table~\ref{table:observing_log} is an observing log. For each epoch, it gives the instrument and its central wavelength. The last column provides the FWHM (in milliarcseconds mas) of a 2D Gaussian fit to the FLUX image of the SPHERE data, and to the core of the PSF of the HiCIAO observations from \citet{mayama2012}. 

\begin{table}
\caption{Observing Log. }
\label{table:observing_log}
\centering   
\begin{tabular}{|c|c|c|c|}
\hline
\hline
\textbf{Epoch}& \textbf{Instrument}&\textbf{Central $\lambda$}[$\mu$m] & \textbf{FWHM [mas$\times$mas]}\\
\hline
Apr.\,11/2012&HiCiAO&1.6 & 70$\times$70\\
Jun.\,10/2015&SPHERE/ZIMPOL&0.626&40$\times$33 \\
Jun.\,30/2016&SPHERE/IRDIS&1.625&56$\times$55\\
Aug.\,13/2017&SPHERE/IRDIS&1.245&40$\times$46\\
Aug.\,14/2017&SPHERE/IRDIS&1.245&39$\times$45\\
Aug.\,17/2017&SPHERE/IRDIS&1.245&40$\times$46\\
Aug.\,18/2017&SPHERE/IRDIS&1.245&43$\times$44\\
Aug.\,22/2017&SPHERE/IRDIS&1.245&40$\times$40\\
Sep.\,04/2017&SPHERE/IRDIS&1.245& --- \\
Sep.\,06/2017&SPHERE/IRDIS&1.245&40$\times$46\\
Sep.\,16/2017&SPHERE/IRDIS&1.245&42$\times$44\\
\hline
\hline
\end{tabular} 
\tablecomments{The Sep.\,04/2017 epoch is omitted because of its FLUX image has a too low signal-to-noise ratio to be fitted. The HiCIAO PSF was not fitted, we report the values given by \citet{mayama2012}.} 
\end{table}

\end{document}